\documentclass[aps,prl,floats,twocolumn,showpacs,superscriptaddress]{revtex4-1}
\usepackage[dvips]{graphicx}
\usepackage{epsfig}
\usepackage{subfigure}
\usepackage{amsmath}
\usepackage{epsfig}
\usepackage{color}

\newcommand{\third}{\mbox{$\frac{1}{3}$}}

\newcommand{\bs}{\mbox{\protect\boldmath $s$}}

\begin{document}

\title{Mean Field Theory For Non-Equilibrium Network Reconstruction}

\author{Yasser Roudi}
\affiliation{NORDITA, Stockholm, Sweden}
%\affiliation{Kavli Institute for Systems Neuroscience, NTNU, Norway}
\author{John Hertz}
\affiliation{NORDITA, Stockholm, Sweden}
\affiliation{The Niels Bohr Institute, Copenhagen, Denmark}

\begin{abstract}
There has been recent progress on inferring the structure of 
interactions in complex networks when they are in stationary states satisfying 
detailed balance, but little has been done for non-equilibrium systems.  
Here we introduce an approach to this problem, considering, as an example, 
the question of recovering the interactions in an asymmetrically-coupled,
synchronously-updated Sherrington-Kirkpatrick model. We derive an exact iterative inversion 
algorithm and develop efficient approximations based on dynamical 
mean-field and Thouless-Anderson-Palmer equations that express the interactions in terms 
of equal-time and one time step-delayed correlation functions.  
\end{abstract}
\maketitle
%\section{Introduction}
{\bf Introduction.}---
Finding the connectivity in complex networks is crucial
for understanding how they operate. 
Gene and multi-electrode microarrays have recently made the type of data required
for this purpose available. What is needed now is appropriate 
theoretical tools for analyzing these data and extracting the connectivity.

In much recent work on this subject 
\cite{Schneidman05,*Shlens06,*Tkacik09,WeigtPNAS2009,*Lezonetal06,Roudi09-2}, 
the problem has been posed as that of inferring the parameters of 
a stationary Gibbs distribution modeling the system. 
While satisfied in many applications, the assumption of Gibbs 
equilibrium is unlikely to hold for many biological systems since they
are usually driven by time-dependent 
external fields, their interactions may not satisfy detailed balance, or they 
may only be observed while the transients dominate the dynamics. Applying 
the equilibrium approach to such cases usually yields effective 
interactions that do not bear an obvious relationship to the 
real ones \cite{Roudi09-2}. Kinetic and nonequilibrium models provide
a much richer platform for studying such systems
\cite{Marre09,Cocco09,Truccolo05}. 

Whereas for equilibrium models the development of systematic 
mean field inference methods
\cite{Kappen98,*Tanaka98} has led to great
practical and conceptual advancements, a mean field theory
for nonequilibrium network reconstruction is still lacking. In this paper, we 
show how a mean field theory for inference can also
be developed for a nonequilibrium system. 
We consider this problem for 
a particular simple nonequilibrium model: a kinetic Ising model with random 
asymmetric interactions ($J_{ji}$ independent of $J_{ij}$),
in an external field which may be time-dependent. 
This is a discrete-time, synchronously updated
model composed of $N$ spins $s_i = \pm 1$ with transition probability
\begin{equation}
\Pr(\bs(t+1)|\bs(t))=\prod_i \frac{\exp[s_i(t+1)\theta_i(t)]}{2\cosh(\theta_i(t))}
\label{DynSK}
\end{equation}
where $\theta_i(t)=h_i(t)+\sum_j J_{ij} s_j(t)$.
The couplings $J_{ij}$ are independent Gaussian variables with 
variance $g^2/N$. This model can be readily applied to time-binned neural 
data, where $t$ labels the bins, and $s_i(t)=\pm 1$ represents a spike or 
no spike by neuron $i$ in bin $t$ \cite{Schneidman05,*Shlens06}. The temperature 
has been set equal to $1$, since any effects of changing the
temperature can be realized by changing the coupling 
parameter $g$ and the field strengths. Even for time-independent field 
and in a steady state, this system is not in a Gibbs 
equilibrium \cite{Coolen00}. However, we show that, like its equilibrium 
counterpart, the nonequilibrium inverse problem for this 
model can be solved using a gradient descent method
and also via systematic approximate inferences derived using dynamical
versions of naive mean-field (nMF) and Thouless-Anderson-Palmer (TAP) equations.
We show that for both the stationary and nonstationary systems 
these methods provide efficient reconstruction of 
interactions. We also analytically quantify their errors.

%\section{Exact, nMF and TAP learning rules}
{\bf Exact, nMF and TAP learning rules.}---
Suppose that we have 
observed $R$ realizations of duration $L$ time steps of 
the process in \eqref{DynSK}. We denote the observed state of the system at time $t$ of 
realization $r$ by $\bs^r(t)=\{s^r_1(t),\cdots, s^r_N(t)\}$.
To find the couplings and external fields, we maximize 
the likelihood of the observed 
states under the model \eqref{DynSK}.
%\begin{eqnarray}
%{\cal L}(\bh,{\sf J})=\sum_{t,r,i}&\big[& h_i s_i^r(t+1) 
%+\sum_j J_{ij} s_i^r(t+1)s_j^r(t)\nonumber\\
%&-&\log 2\cosh(h_i(t)+\sum_j J_{ij} s^r_j(t))\big].	
%\label{loglike}
%\end{eqnarray}
This maximization can be done using an iterative algorithm, analogous 
to Boltzmann learning for the equilibrium model: starting from an 
initial set of couplings and fields, one adjusts them iteratively
by steps of sizes $\delta h_i=\eta_{h} \frac{\partial {\cal L}}{\partial h_i}$ 
and $\delta J_{ij}=\eta_J \frac{\partial {\cal L}}{\partial J_{ij}}$, ${\cal L}$ 
being the log-likelihood. The learning steps thus are
\begin{subequations}
\begin{align}
&\delta h_i(t) =
\eta_{h}\big\{\langle s_i(t+1)\rangle_r-\langle \tanh[\theta_i(t))] \rangle_r ]\big\} \label{hstep}\\
&\delta J_{ij}=       
\eta_{J}\big\{\langle s_i(t+1) s_j(t)\rangle-\langle \tanh[\theta_i(t) ]s_j(t)\rangle\big \}\label{Jstep}
\end{align}
\label{LR}
\end{subequations}
where $\eta_h$ and $\eta_J$ are learning rates. Here and in what follows
$\langle \cdots \rangle_r$, $\langle \cdots \rangle$
represent averaging over repeats, and both repeats and time, respectively.
An overline, instead, will indicate averaging over the spins. One can think of
Eq.\ \eqref{Jstep} as performing a logistic regression
to explain one-step separated correlations. This is similar to
what is proposed in \cite{Ravikumar10} as an approximation
for inferring the connectivity in an equilibrium Ising model. 

Since performing  the steps in this algorithm does not require Monte Carlo runs, 
it is faster than the equilibrium Boltzmann learning. 
However, two factors still make this algorithm slow for 
large systems and/or data sets, warranting the development of fast approximations.
First, \eqref{LR} is still an iterative algorithm which could take 
a long time to converge if not provided with a good initial condition
and learning rates. Second, at each step the averages
on the right hand side of \eqref{LR} should be calculated from 
the data {\em de novo}, given the adjusted parameters. 

Two fast approximations, nMF and TAP learning rules, are derived and
studied below. To implement them in the stationary case, 
one first uses the data to calculate the 
one-step delayed and equal time correlations,
$D_{ij} = \langle \delta s_i(t+1) \delta s_j(t)\rangle$ and
$C_{ij} = \langle \delta s_i(t) \delta s_j(t)\rangle$, where
$m_i = \langle s_i \rangle$ and $\delta s_i=s_i-m_i$. The 
approximations are
\begin{equation}
{\sf J}^{\rm nMF/TAP}={\sf A^{nMF/TAP}}^{-1} {\sf D} {\sf C}^{-1} \label{JMF}
\end{equation}
where $A^{\rm nMF}_{ij}=(1-m_i^2) \delta_{ij}$, 
$A^{\rm TAP}_{ij}=A^{\rm nMF}_{ij} (1-F_i)$ and
$F_i$ is the root of the cubic equation \eqref{cubiceqn} below.
In the nonstationary case too, similar learning rules can be derived
as shown later in the paper.

%\section{Derivation of nMF and TAP inversion}
{\bf Derivation of nMF and TAP inversion.}---
For simplicity, we consider first the stationary case, for which 
the sequence index $r$ is superfluous, as averaging over time and 
repeats would be equivalent.
We start with the maximum 
likelihood conditions, i.e. $\delta h_i=\delta J_{ij}=0$ in \eqref{LR}. 
Using the nMF equations $m_i = \tanh (h_i + \sum_j J_{ik}^{\rm nMF}m_k)$, and
writing the $s_i$ in \eqref{LR} as $m_i+\delta s_i$, we expand
the tanh in the $\delta s_i$. The first nonzero term gives
\begin{equation}
\langle \delta s_i(t+1) \delta s_j(t)\rangle = (1-m_i^2)\sum_k J_{ik}^{\rm nMF}\langle \delta s_k(t) \delta s_j(t)\rangle.
\label{MFresult}
\end{equation}
which can be written as \eqref{JMF} for the nMF case.

To get the TAP inversion formula, we start instead by assuming that 
the $m_i$ satisfy the TAP equations 
$m_i = \tanh [h_i + \sum_k J_{ik}^{\rm TAP}m_k -m_i\sum_k (J^{\rm TAP})_{ik}^2 (1-m_k^2)]$,
which take into account
the Onsager reaction term. Kappen and Spanjers \cite{Kappen00} proved that the TAP equations, 
although usually derived for the equilibrium (symmetric-${\sf J}$) SK model,
also hold for the {\em asynchronously updated}, asymmetric-$\sf J$ 
model in a stationary state. We have verified that they are also valid 
in our synchronously-updated model \cite{Roudi10-2}. We again 
write $s_i = m_i + \delta s_i$, expanding the tanh to third order in 
powers of $\sum_kJ_{ik}^{\rm TAP}\delta s_k + m_i\sum_k(J^{\rm TAP})_{ik}^2(1-m_k^2)$.
Keeping terms up to order $g^3$ leads 
to  ${\sf D} = {\sf A}^{\rm TAP}{\sf J}^{\rm TAP}{\sf C}$ , where
\begin{equation}
A_{ij}^{\rm TAP} = A^{\rm nMF}_{ij} [1 - (1-m_i^2)\sum_l (J^{\rm TAP})_{il}^2 (1-m_l^2)].
\label{ATAP}
\end{equation}
These equations cannot be solved directly as in the nMF case because ${\sf A}^{\rm TAP}$ 
depends on ${\sf J}^{\rm TAP}$. However, one can derive a cubic equation 
for the quantities $F_i = (1-m_i^2)\sum_l (J^{\rm TAP})_{il}^2 (1-m_l^2)$:
\begin{equation}
F_i (1-F_i)^2 = (1-m_i^2)\sum_j(J^{\rm nMF})_{ij}^2(1-m_j^2).
\label{cubiceqn}
\end{equation}
This determines $A^{\rm TAP}_{ij} =A^{\rm nMF}(1-F_i)$, yielding
\eqref{JMF} for the TAP case. The relevant root
of \eqref{cubiceqn} is the smallest one (the one approaching zero as $g\to 0$).
This root cannot exceed $1/3$, restricting this technique 
to weak couplings.

For both nMF and TAP reconstruction, the external fields $h_i$ can also be found by solving the 
respective magnetization equations after the $J_{ij}$ have been obtained, 
just as in the equilibrium problem \cite{Kappen98,Tanaka98}.  
%\section{Performance of the algorithms}

{\bf Performance of the algorithms.}---
We have verified that the algorithm \eqref{LR} recovers the couplings of an 
asymmetric SK model exactly in the limit of $L \to \infty$, for 
a wide range of coupling strengths $g$, external fields and system sizes. The 
mean square error, $\epsilon_{\rm exact}$, is in general proportional to $1/L$, and in the 
weak-coupling limit a quadratic expansion of log-likelihood yields 
\begin{equation}
\epsilon_{\rm exact} = \overline{\delta J_{ij}^2} \equiv \overline{(J_{ij} - J_{ij}^{\rm true})^2} = \frac{1}{(1-m_i^2)L},				\label{exacthighT}
\end{equation}
where $J_{ij}(J^{\rm true}_{ij})$ are the inferred (true) couplings.

\begin{figure}[hbtp]
\subfigure{\includegraphics[height=1.3in, width=1.65in]{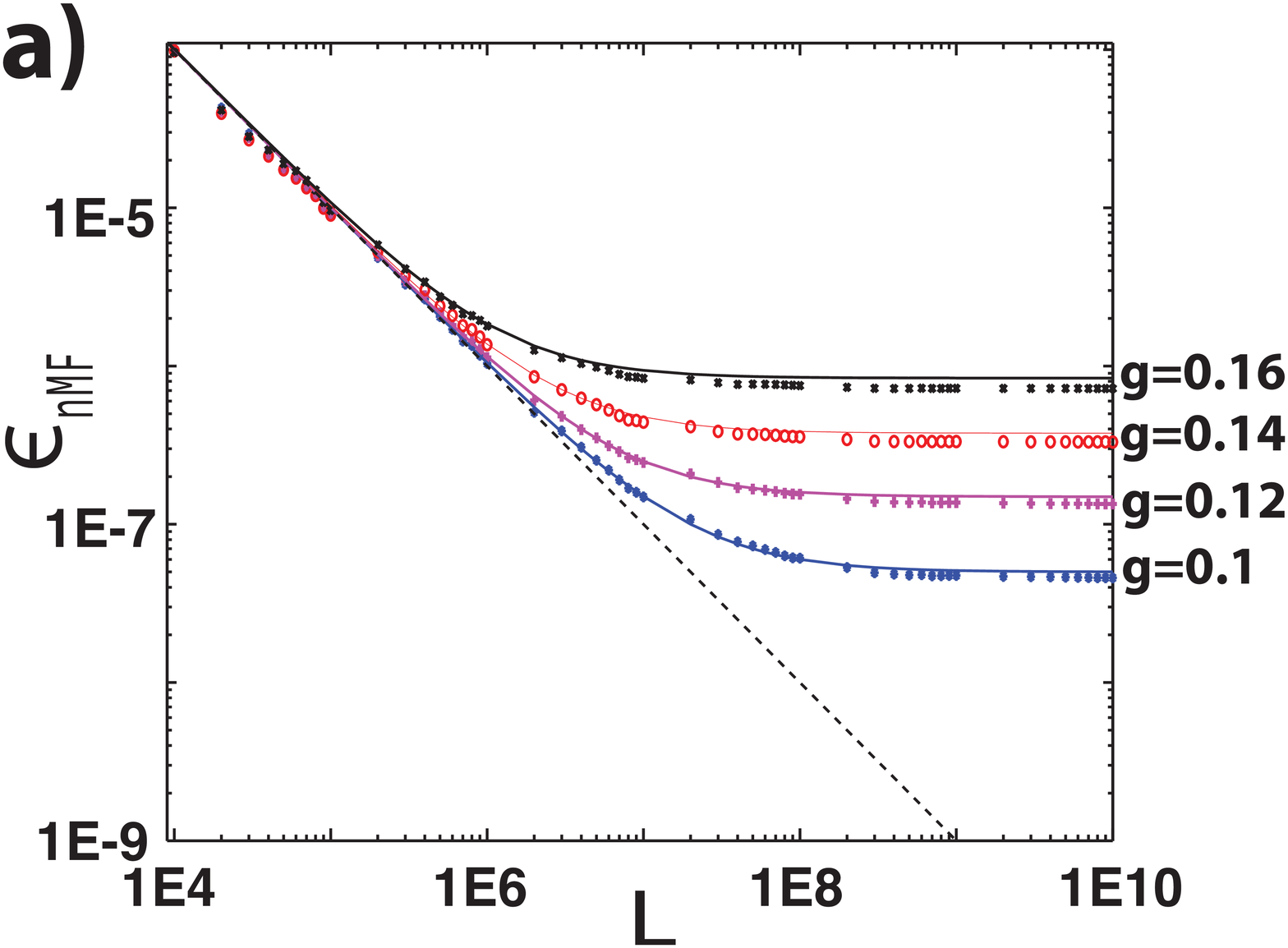}}
\subfigure{\includegraphics[height=1.3in, width=1.65in]{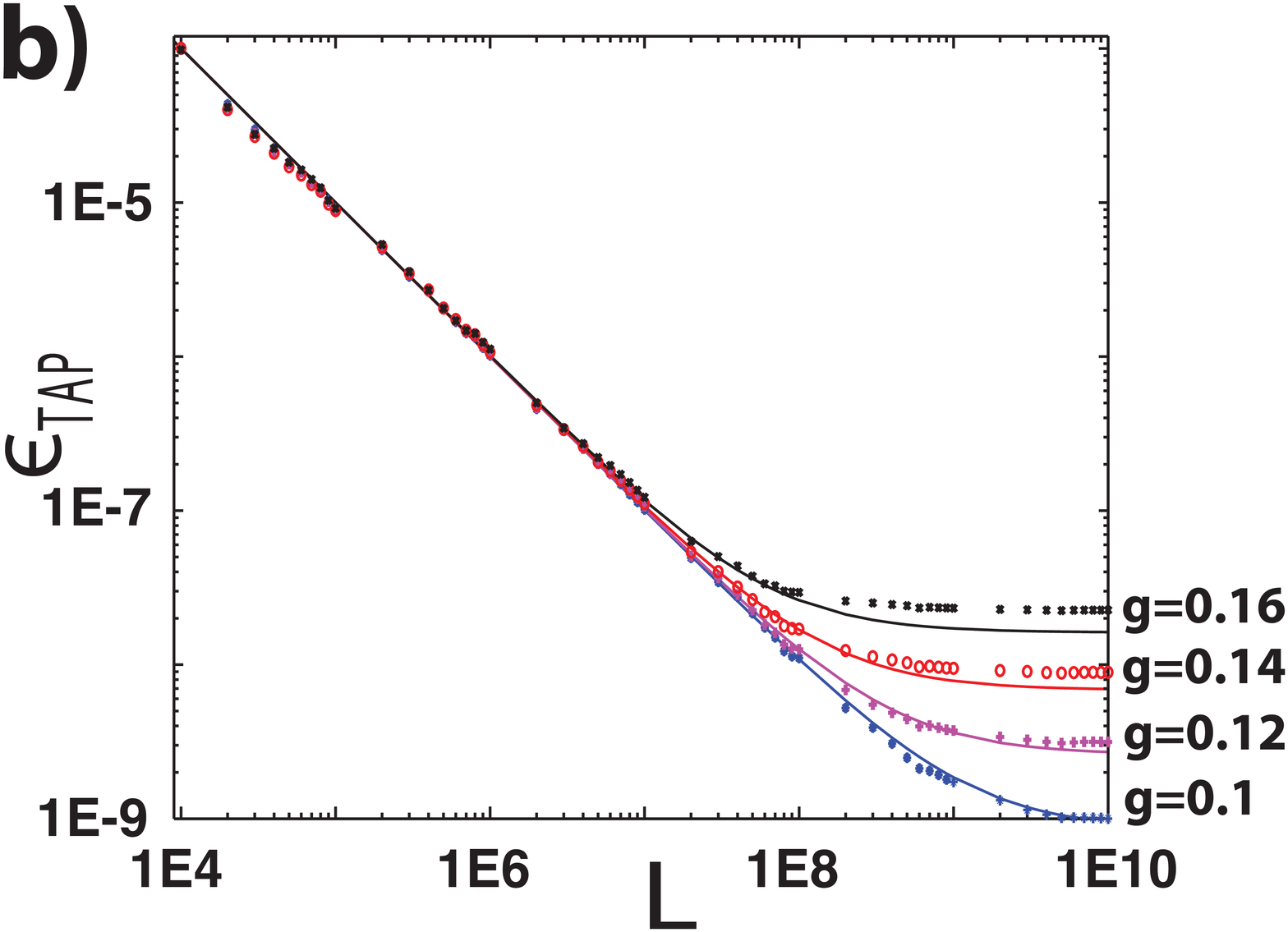}}
\caption{(Color online) Performance of the algorithms. Exact and nMF (a).
and the TAP (b) erros are shown vs data length $L$
for $g=0.1$ (blue stars), $0.12$ (magenta crosses), $0.14$ (red circles)
and $0.16$ (black x), all for $N=20$ and zero external field. 
Theoretical predictions are the solid lines.}
\label{Fig1}
\end{figure}

We find that the nMF algorithm leads to an error, $\epsilon_{\rm MF}$, of the 
form $\epsilon_{\rm exact} + \epsilon^{\infty}_{\rm nMF}$, where $\epsilon^{\infty}_{\rm nMF}$ 
is independent of $L$ and proportional to $1/N$. Thus, for data sets of 
length $L\ll L^* = 1/\epsilon^{\infty}_{\rm nMF} \propto N$, nMF does almost 
as well as the exact algorithm. Furthermore, the larger the network, the 
better nMF does. The errors for the exact and nMF algorithms vs $L$ 
are shown in Fig.\ \ref{Fig1}a.  

For weak coupling, we can calculate the asymptotic nMF error, $\epsilon^{\infty}_{\rm nMF}$, 
analytically as follows. We present the zero-field case here for simplicity. 
We expand the tanh in the max-likelihood equation to third order, giving
\begin{equation}
D_{in} = \sum_k J_{ik} \langle s_k s_n \rangle -\third \sum_{klm}J_{ik}J_{il}J_{im}\langle s_k s_l s_m s_n \rangle + \cdots	.		
\label{MFexpand}					
\end{equation}
Correlations here are at equal times, except for $D_{in}$. The dominant 
contributions in the sum over $k,l,m$ are those with $k=l$, $l=m$ and $m=k$. 
Multiplying on the right by $(C^{-1})_{nj}$, summing over $n$ and using 
\eqref{JMF} for nMF, yields
\begin{equation}
J^{\rm nMF}_{ij} = J_{ij}- \sum_k J_{ik}^2 J_{ij},
\label{MFcorrections}
\end{equation}
with corrections of relative order $1/N$. Eq. \eqref{MFcorrections} also 
yields the TAP-approximation couplings found above, showing that the 
TAP reconstruction indeed corrects 
the leading MF errors. To leading order the sum on $k$ is just $g^2$, and 
the asymptotic mean square MF error is
\begin{equation}
\epsilon_{\rm nMF}^{\infty} = \overline{ (J_{ij} - J_{ij}^{\rm nMF})^2 } = \frac{g^6}{N}.    \label{MFmse}
\end{equation}
The solid curves in Fig.\ \ref{Fig1}a are $1/L + g^6/N$; the fit is evidently good.  
As shown in Fig.\ S1 \cite{EPAPS}, nMF 
exhibits a systematic error by underestimating 
the magnitude of the couplings. The factor $1-F_i$ in TAP formula 
corrects for this to relative order $g^2$. Thus, when one is 
interested only in the presence or absence of connections,
there would be little difference between nMF and TAP.

The error for the TAP reconstruction
is much lower than that of the nMF 
one and reaches its minimum at much larger $L$: for $N=20$ and the coupling 
strengths we studied, we had to 
go to $L\sim 10^9$ to see the error flatten (Fig.\ \ref{Fig1}b). To calculate the 
asymptotic reconstruction error for TAP, we expand the tanh
to 5th order and proceed to evaluate the averages as we did for nMF.
The nMF error terms analyzed above are compensated
for by the TAP equations, as $N \to \infty$, leading to an 
asymptotic $\epsilon_{\rm TAP}^{\infty} = 4g^{10}/N$.  
For $N\gg 1/g^2$ this is the leading term in the
asymptotic TAP error. Outside this regime, a
finite-size effect should also be
taken into account. This is because in making that TAP 
correction, the term in \eqref{MFexpand} with $k=l=m$ has been 
counted three times in obtaining \eqref{MFcorrections} instead 
of once. The mean square error that results from this overcounting is
%\begin{equation}
$(2/3)^2 \overline{J_{ij}^6} =  (20g^6)/(3N^3)$ and should be 
added to the $4g^{10}/N$ term.
%\label{fsc}
%\end{equation}

%\section{Non-stationary case}
{\bf Non-stationary case.}---
The magnetizations, $m_i(t)=\langle s^r_i(t)\rangle_r$, are now time-dependent and,
for nMF, solve
\begin{equation}
m_i(t+1) = \tanh [h_i(t)  + \sum_jJ_{ij}^{\rm nMF}m_j(t)].				
\label{MFdyn}
\end{equation}
We have also proved \cite{Roudi10-2} 
that the TAP equations hold even in a 
nonstationary state, in the form
\begin{eqnarray}
m_i(t+1) &=& \tanh [h_i(t)  + \sum_jJ_{ij}^{\rm TAP}m_j(t)					\nonumber \\
 &-& m_i(t+1)\sum_j(J^{\rm TAP})_{ij}^2(1-m_j^2(t))].	    \label{dyn}
\label{TAPdyn}
\end{eqnarray}
Thus, we can extend both our inversion 
algorithms to nonstationary systems, as we show in the following.

We start by defining time-dependent 
correlation matrices $D_{ij}(t) \equiv \langle \delta s_i^r(t+1) \delta s_j^r(t) \rangle_r$ 
and $C_{ij}(t) \equiv \langle \delta s_i^r(t) \delta s_j^r(t) \rangle_r$. 
For nMF, using the same procedure that lead to \eqref{MFresult}, we find
\begin{equation}
\langle D_{ij}(t)\rangle_t  = \sum_k J_{ik}^{\rm nMF} \langle (1-m_i^2(t+1))C_{kj}(t) \rangle_t . 
\label{invdynMF}
\end{equation}
One can still solve for $\sf J$ by simple matrix algebra:
\begin{equation}
J_{ij}^{\rm nMF} = \sum_k \langle D_{ik}(t)\rangle_t [({\sf B}^{(i)})^{-1}]_{kj},		
\label{dyninvsoln}
\end{equation}
where $B_{kj}^{(i)} =  \langle (1-m_i^2(t+1))C_{kj}(t) \rangle_t$. 	
%\begin{equation}
%B_{kj}^{(i)} =  \langle (1-m_i^2(t+1))C_{kj}(t) \rangle_t 						
%\label{BMF}
%\end{equation}
The problem is more complex than the stationary one only because 
one has to invert a different matrix ${\sf B}^{(i)}$ for each $i$.  

For TAP, analogously to the stationary case, the ${\sf B}^{(i)}$ 
acquire an extra factor inside the time average:
\begin{subequations}
\begin{align}
&B_{kj}^{(i)} =  \langle (1-m_i^2(t+1))(1-F_i(t))C_{kj}(t)  \rangle_t ,			
\label{BTAP}\\
&F_i(t) = (1-m_i^2(t+1))\sum_l (J^{\rm TAP})_{il}^2 (1-m_l^2(t)).		
\label{FdynTAP}
\end{align}
\label{dynTAP}
\end{subequations}			
Exact TAP inversion requires iterative solution of  
\eqref{dyninvsoln}, with $J^{\rm TAP}_{ij}$ instead
of $J^{\rm nMF}_{ij}$, together with \eqref{dynTAP}.  
We have found, however, that effective reconstruction is still 
possible under the simplifying approximation that
$F_i(t)$ in Eq.\ \eqref{BTAP} can be represented by its 
temporal mean.
In this case, $F_i \equiv \langle F_i(t) \rangle_t $ solves the cubic equation  
\begin{equation}
F_i(1-F_i)^2 = \sum_j (J^{\rm nMF})_{ij}^2 \langle (1-m_i^2(t+1))(1-m_j^2(t)) \rangle_t.\nonumber
%\label{approxdyncubic}
\end{equation}
Solving it and using it in Eq. \eqref{BTAP}, one can calculate
$J_{ij}^{\rm TAP} = J_{ij}^{\rm nMF}/(1-F_i)$.
Similar to the stationary case, after inferring the couplings, one can use
the forward dynamical nMF and TAP equations Eqns.\ \eqref{MFdyn} and \eqref{TAPdyn}
to infer the time-varying external field.
\begin{figure}[hbtp]
  \subfigure{\includegraphics[height=1.1in, width=1.5in]{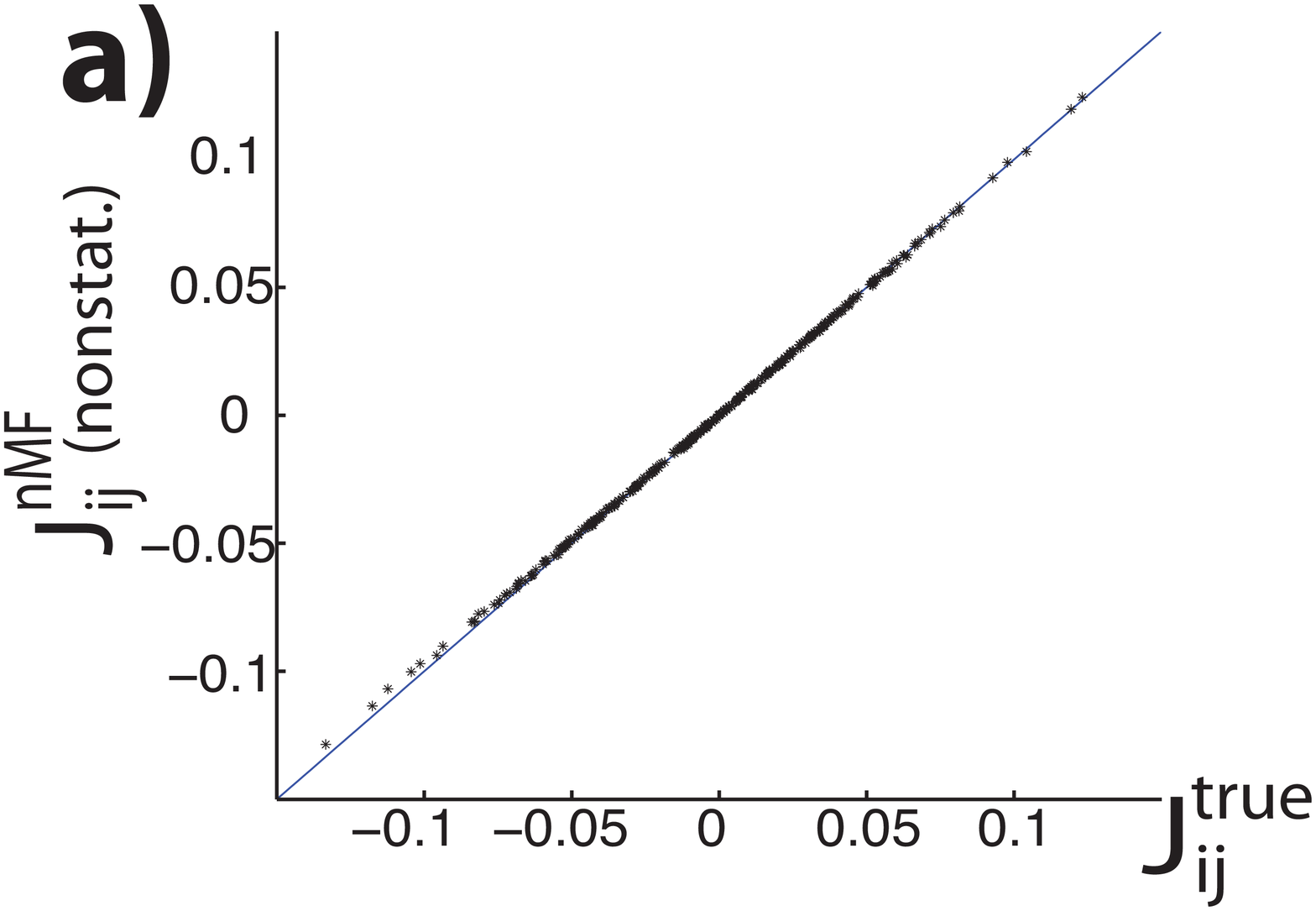}}
  \subfigure{\includegraphics[height=1.1in, width=1.5in]{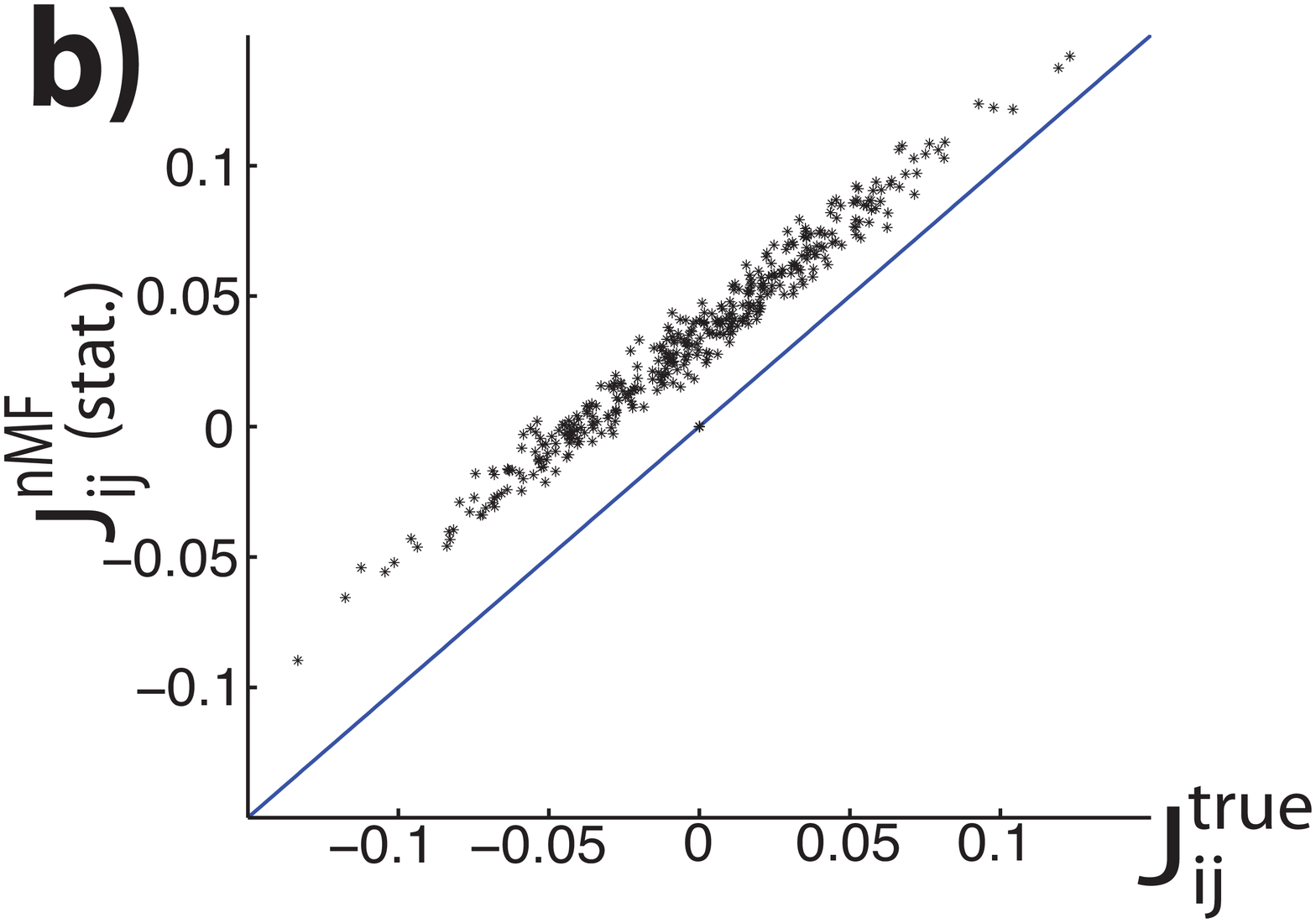}}\\
  \subfigure{\includegraphics[height=0.9in, width=2.3in]{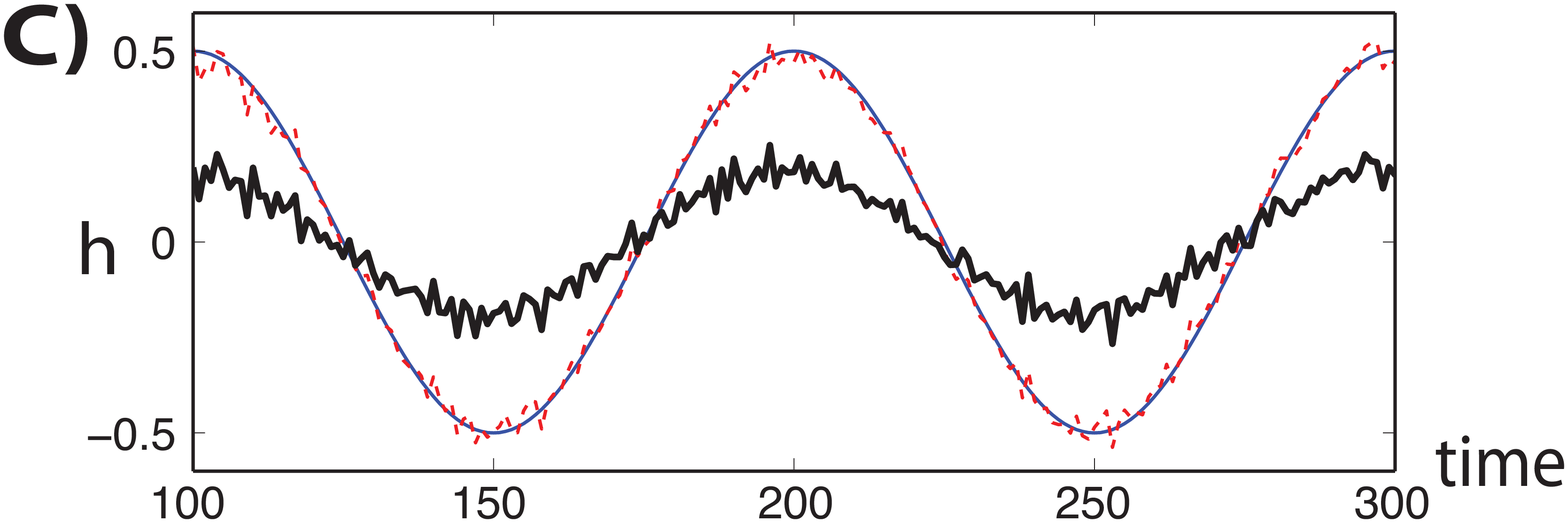}}
  \caption{(Color online) Inference in the nonstationary case. (a) Couplings of a network of $N=20$
 driven by a sinusoidal external field inferred using the nonstationary nMF, and (b)
 the stationary nMF.
 (c) Two periods of the external field (thin blue full curve) and its reconstruction using
 the nonstationary nMF couplings (red dashed curve) and stationary nMF (thick black full curve).}
\label{Fig3}
\end{figure}
The result of reconstructing the couplings of a
network driven by a common sinusoidal external field 
to all spins is shown in Fig.\ \ref{Fig3}. Fig.\ \ref{Fig3}a shows how well 
the couplings are inferred 
by nonstationary MF using $L=10^5$ and $R=100$. Nonstationary TAP couplings (not shown)
have a lower mean squared error:  $6.7\times 10^{-7}$ versus $10^{-6}$ for nMF.
In Fig.\ \ref{Fig3}b, we also plot the couplings inferred using 
{\em stationary} nMF inversion for each of the $100$ repeats and averaging over them.
Not surprisingly, the stationary nMF performs poorly on this nonstationary data.
Importantly, there is a systematic overestimation 
of the couplings in this case, because
the stationary method accounts for correlations induced by 
the common, time-varying external field through adjusting the couplings. 
Correspondingly, if one uses the couplings inferred by
stationary nMF in \eqref{MFdyn} to infer $h_i(t)$,
the amplitude of this field is underestimated, while the use of
nonstationary nMF couplings yields a very good reconstruction 
of $h_i(t)$; see Fig.\ \ref{Fig3}c.

%\section{Discussion} 
{\bf Discussion.}---   
We have shown how to infer interactions in a simple but nontrivial 
nonequilibrium system: a kinetic Ising model with random and 
potentially asymmetric interactions.  The model is the
maximum entropy model {\em for each time step}, 
given mean magnetizations and one step separated correlations.
We have described 
both an exact iterative algorithm and two approximate ones, based 
on dynamical nMF and TAP equations, which are correct 
up to corrections of order $1/N$. We calculated analytically the 
errors of these approximations for weak coupling. The 
method shows particular promise when applied to nonstationary 
states, where it separates true interactions from the apparent ones 
found by applying a stationary theory to a nonstationary state.  
\begin{figure}[hbtp]
  \subfigure{\includegraphics[height=1.2in, width=1.4in]{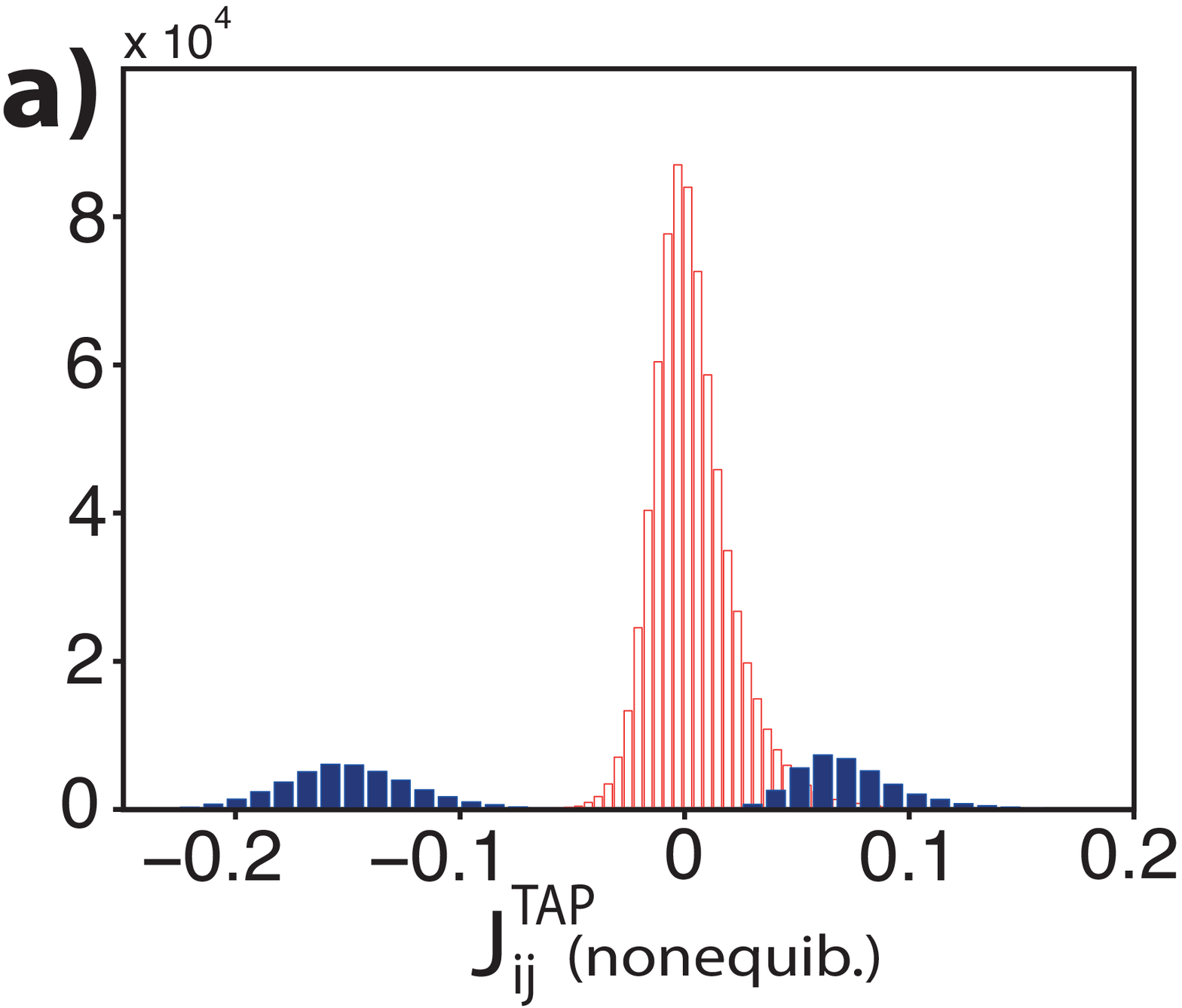}}
  \subfigure{\includegraphics[height=1.2in, width=1.4in]{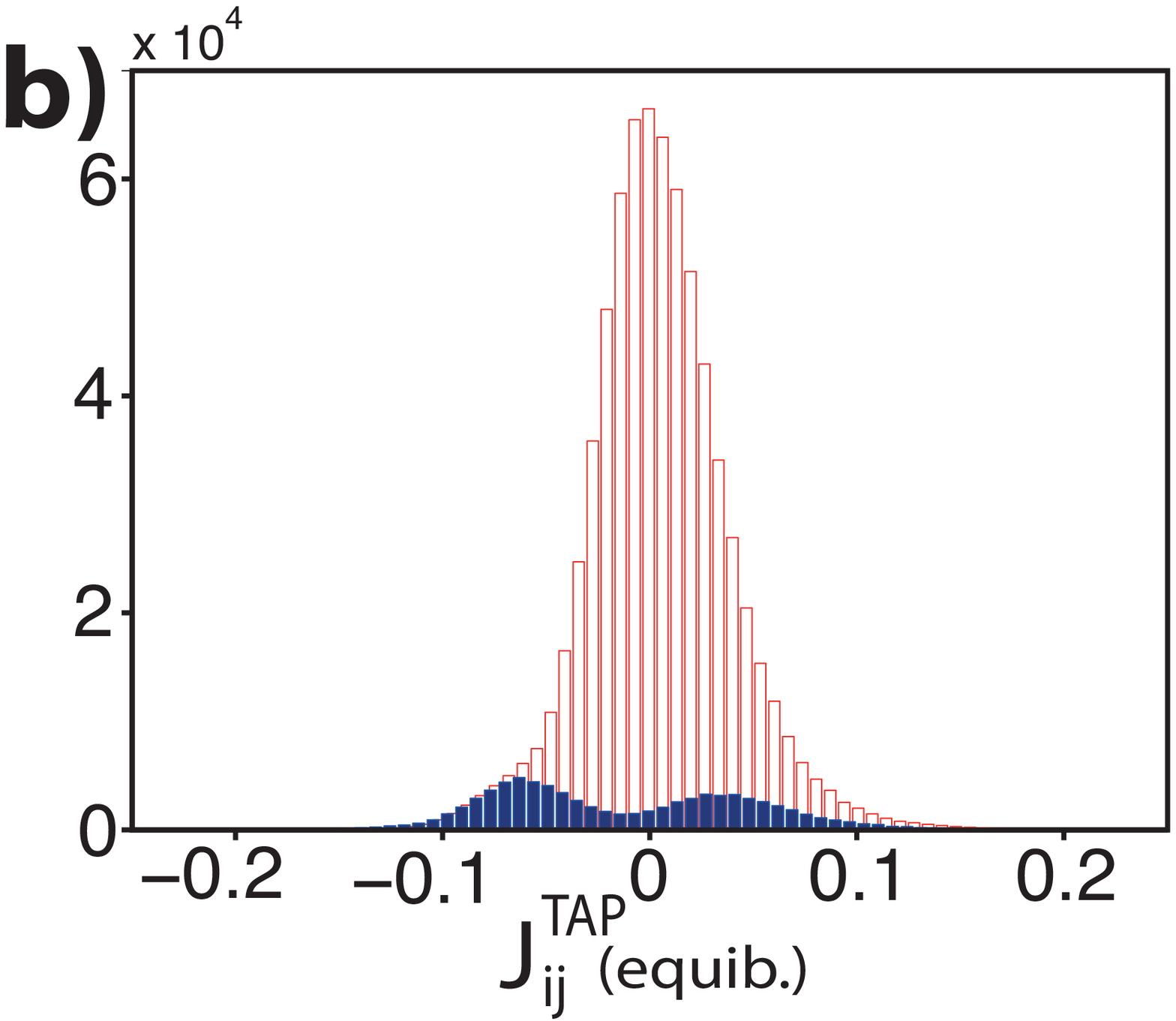}}
\caption{(Color online) Finding connections in a cortical network model. 
(a) The histogram of the couplings inferred using the stationary nonequilibrium 
TAP for pairs of neurons that were connected (blue full bars), 
and those that were not (red empty bars).
The separation between the histograms shows that one 
can use the TAP approximation to separate connected and disconnected pairs.
(b) same as (a) for equilibrium TAP.}
\label{Fig4}
\end{figure}

A kinetic Ising model will show an intrinsic
error when applied to data from a different kind of system.
However, even when applied to data from a
realistic network, the simple approximate learning rules
developed here identify the connections much better
than their equilibrium counterparts. Fig. \ \ref{Fig4} 
shows the distribution of couplings found
by applying the nonequilibrium TAP learning to 
data from a simulated model cortical column with inhibitory and excitatory neurons \cite{Hertz10}.
The connections in the model were dilute with $10\%$ probability of connection.
When there is no synapse from neuron $j$ to $i$, the inferred $J_{ij}$
follows a zero mean distribution, while if there is an excitatory/inhibitory
synapse, it follows a positive/negative mean distribution,
well separated from the first one.
One can thus easily use the distribution of inferred couplings to infer the
presence, absence and sign of the connections;
see \cite{EPAPS} and \cite{HertzCNS10}. On the contrary, the resulting distributions
are completely overlapping when and equilibrium TAP learning is used.
When using a model like \eqref{DynSK} to infer connectivity
in a system with a different dynamics, or when
faced with data limitation, including prior knowledge about the 
network could be very beneficial. In particular, taking
into account sparsity of the connections via a $l$-1 regularizer
added to the likelihood has been shown to be very useful \cite{Ravikumar10}. 
It is easy to show that adding an $l$-1 
regularizer to the likelihood of the data under \eqref{DynSK} 
would modify \eqref{JMF} by adding a
term proportional to ${{\sf A}^{\rm nMF/TAP}}^{-1} {\rm sgn}({\sf J}) {\sf C}^{-1}$ to the right hand
side. How this improves inferring connections in biological networks
will be discussed elsewhere.

A simple extension of  \eqref{DynSK} is its continuous
time version. As shown in \cite{Zeng10}, for this
model, too, a mean field theory can be developed using the approach presented here.
In other recent kinetic approaches to problems like this,  
the equilibrium maximum-entropy approach 
\cite{Schneidman05,Shlens06} is extended to include 
non-equal-time correlations \cite{Marre09} and an approximate 
scheme for fitting an integrate-and-fire network to data was developed in
\cite{Cocco09}. There 
has also been work \cite{Truccolo05}, closely connected to
\eqref{DynSK}, in which $s_i(t+1)$ depends on linear 
combinations of $h(t')$ and $s(t')$, for $t' \le t$.   
Given the advantage of these nonequilibrium models 
over the equilibrium ones for describing spike train statistics, a mean-field theory for inferring
their parameters would be of great theoretical and practical benefit.
For such models, we expect
that it will be possible to use the techqniues in \cite{Kappen00} or \cite{Biroli99,Coolen00}
to derive dynamical nMF and TAP equations. Employing the approach 
developed here one can then build approximate mean field inversion techniques 
based on these equations.  
\end{document}